\documentclass[conference]{IEEEtran}
\IEEEoverridecommandlockouts

\usepackage{cite}
\usepackage{amsmath,amssymb,amsfonts}
\usepackage{algorithmic}
\usepackage{graphicx}
\usepackage{textcomp}
\usepackage{xcolor}
\usepackage{xspace}
\usepackage{subcaption}
\usepackage{multirow}
\usepackage{siunitx}
\usepackage{wasysym}
\usepackage{booktabs}
\usepackage{tabularx}

\newcommand{\todo}[1]{\textcolor{red}{TODO: #1}}
\newcommand{\prh}[1]{\noindent{\textbf{#1}}}

\newcommand{\transpiler}[1]{Delorean\xspace{}}
\newcommand{\tool}[0]{\textsc{\textsc{VAPy}}\xspace{}}

\newcommand{\txs}[1]{\textsf{\small #1}}
\newcommand{\txss}[1]{\textsf{\scriptsize #1}}

\newcommand{\circulax}{Circulax\xspace{}}
\newcommand{\jax}{JAX\xspace{}}
\newcommand{\jit}{JIT\xspace{}}
\newcommand{\osdi}{OSDI\xspace{}}
\newcommand{\openvaf}{OpenVAF\xspace{}}

\newcommand{\veriloga}{\textsc{VA}}

\newcommand{\swatch}[1]{\textcolor{#1}{$\blacksquare$}}
\def\BibTeX{{\rm B\kern-.05em{\sc i\kern-.025em b}\kern-.08em
    T\kern-.1667em\lower.7ex\hbox{E}\kern-.125emX}}
\begin{document}

\title{\tool{}: Circuit Optimization with Backpropagation
}

\author{
\IEEEauthorblockN{
    Anthony Yang\textsuperscript{1},
    Ron Dubinsky\textsuperscript{2},
    Yu-Neng Wang\textsuperscript{1},
    Sara Achour\textsuperscript{1,\textdagger}
}
\IEEEauthorblockA{
    \textsuperscript{1}Stanford University,
    Stanford, CA, USA\\
    \textsuperscript{2}Independent Researcher,
    Redmond, WA, USA\\
    \textsuperscript{\textdagger}Corresponding Author \\
    \{anthyang, wynwyn, sachour\}@stanford.edu;
    rondubinsky1@gmail.com\\
}
}

\maketitle

\begin{abstract}
Parameter sensitivities of analog circuits enable the use of gradient-based design optimization methods and unlock a range of sensitivity analyses. However, it remains a challenge to compute exact design parameter sensitivities for circuits that use PDKs, which use Verilog-A device models comprised of thousands of lines of code. Finite difference methods, which estimate gradients through simulation, require $N+1$ to $2N$ simulations for each gradient depending on the method, which can become prohibitively expensive when many parameters are optimized. We present \tool{}, a Verilog-A to \jax{} transpiler that compiles Verilog-A device models into autodifferentiable \jax{} code, enabling optimization with \circulax{}, a \jax{}-based circuit simulator. \tool{} employs taint analysis to ensure only the necessary control-flow is rewritten with \jax{}-compatible calls, reducing control-flow overheads. We use \tool{} and \circulax{} to optimize three analog circuits implemented in the SkyWater SKY130 PDK, which is built on the BSIM4 device model. \tool{}-generated model gradients closely match finite-difference gradients, achieving comparable design quality in over $\mathbf{40\times}$ fewer circuit solves and up to 40\% lower loss than Bayesian optimization.


\end{abstract}

        


\section{Introduction}

Automated design optimization of analog mixed-signal circuits remains a challenge for EDA tools. Design optimization involves optimizing circuit parameters, such as transistor widths, to maximize some figure of merit computed from circuit simulation results. Parameter sensitivity, or gradient, information is valuable for such problems since it can be used to optimize the circuit with gradient-based methods and is independently useful for other use cases, such as yield analysis where practitioners are interested in the sensitivity of their figure of merit to a particular design parameter.

Currently, it is challenging to compute the design parameter sensitivities for circuits comprised of PDK-provided transistor models due to the complexity of these models. Analytic circuit-native methods for computing parameter sensitivities, such as circuit adjoint methods, only support simple (e.g., level 1) transistor models and often require manual effort~\cite{director1969generalized, visweswariah2000noise, hu2020adjoint, li2023circuit, lyu2024bayesianwithsensitivity}. Finite difference methods (FD), which approximate gradients by perturbing individual parameters, are able to handle complex models but require $N+1$ or $2N$ circuit simulations per gradient depending on the method. Therefore, for optimization problems involving many parameters, FD methods scale poorly.

This work focuses on leveraging automatic differentiation (AD) to compute parameter sensitivities, a technique commonly used for analytically deriving gradients from ML models. We present \tool{}, a transpiler that compiles PDK-provided device models to Python and then uses AD to directly compute parameter sensitivities from a single circuit simulation. 

Using \tool{}, we are able to automatically generate Python-based device models for the SkyWater SKY130 PDK~\cite{sky130}, and autodifferentiate circuits implemented in SKY130 with respect to their design parameters. \tool{} is able to autodifferentiate the BSIM4 device model used in SKY130, which is currently not supported by Xyce, a flagship C++ simulator that offers support for AD of models over a subset of the Verilog-A language syntax~\cite{xyce}.

With \tool{}-derived sensitivity functions, only one simulation is required per gradient, substantially reducing the number of simulations required per gradient compared to FD as the number of parameters increases.  With \tool{}, we are able to optimize SKY130 circuits using gradient descent methods while precisely computing the parameter sensitivities, and identify cases where \tool{} sensitivities are more accurate than FD.

\section{Background}

Analog circuits are described by netlists of parameterized devices, such as transistors, capacitors, resistors, and switches. Their topology defines the physical constraints that govern device behavior. These netlists are commonly specified as SPICE circuits, which are simulated with a number of circuit simulators, including SPECTRE, ngSpice, and Xyce. Typically, device physics models are built into the simulator. Device instances expose design parameters while remaining parameters are specified by the foundry the device is fabricated with. A process design kit (PDK) is a foundry-provided library that parametrizes device models to reflect how devices fabricated using their process will behave. A device model in a PDK may be parameterized differently to model p-type and n-type transistors, threshold voltage, process corners, and allowed operating conditions. These foundry parameters may also depend on device design parameters, such as width and length.

\prh{Recent Advances.} Until relatively recently, physics models for parameterized devices were implemented within simulators in C++. Introduced in 2022 by SemiMod GmbH, the Open Source Device Interface (OSDI) enables independently compiled device models to be loaded by compatible
simulators. Several notable simulators, including ngSpice, adopted this standard. In 2023, SemiMod released the OpenVAF compiler, which compiles Verilog-A device models to binaries through an LLVM IR. Verilog-A is an analog behavioral modeling language that many of the BSIM transistor model families are implemented in. These recent advances in open-source analog EDA tooling enable the development of \tool{}.

\prh{Autodifferentiation and ML.} Machine learning libraries, such as PyTorch and \jax{} use automatic differentiation (AD), which computes exact derivatives of a program by applying the chain rule to its elementary operations, allowing for accurate, non-approximate gradients at instantaneous points and eliminating manual processes.  

Very recently, these frameworks have been implemented into circuit simulators. VAJAX~\cite{vajax} and Circulax~\cite{circulax} are \jax{}-based circuit simulation frameworks that were released in 2026 and 2025, respectively~\cite{jax2018github}. VAJAX is built for GPU acceleration of analog circuit simulation and has AD support limited to device residuals and Jacobians rather than arbitrary parameter gradients.

\section{\tool{} Transpiler}\label{sec:tech:compiler}


The \tool{} transpiler translates \veriloga{} device models to a semantically equivalent Python abstract syntax tree (AST) through source-to-source translation ($\S$~\ref{sec:tech:transpiler}), which is then compiled into an autodifferentiable JAX implementation compatible with Circulax, a Python-based SPICE circuit simulator ($\S$~\ref{sec:tech:circsim}). Section~\ref{sec:tech:devlib} describes how the \tool{}-generated device models are used with Circulax to perform circuit optimization with backpropagation.

\subsection{Verilog-A to Python Transpiler}\label{sec:tech:transpiler}

\tool{}'s transpilation front-end parses and translates a \veriloga{} program into a Python AST. 
\tool{}'s parser and intermediate representations extend the \openvaf{} model compiler, which translates Verilog-A device models to OSDI-compliant binary code. 

\prh{The \osdi{} Standard.} \tool{} compiles \veriloga{} device models into Python representations based on the Open Source Device Interface (\osdi{}) standard. The OSDI standard defines an application binary interface that ensures interoperability with common open-source SPICE simulators. \tool{} extends OpenVAF-reloaded \cite{openvaf-reloaded}, an open-source compiler for Verilog-A based on OpenVAF that generates models conforming to \osdi{} v0.4. This fork improves the OpenVAF compiler for \veriloga{}, which targets version \osdi{} v0.3. 

\prh{Transpilation.} 
\tool{} expects a device model OpenVAF can compile and express as supported scalar equations. \tool{} transpiles the code into a Python representation containing \txs{setup}, \txs{init}, and \txs{eval} functions, resembling the \txs{setup\_model}, \txs{setup\_instance}, and \txs{eval} found in \osdi{}-compliant models. In \tool{}'s abstraction, the \txs{setup} function creates a new device model object $m$ from PDK parameters $\phi$. The $\vec x =$\txs{init}($m$,$\theta$) function instantiates a device instance with instance-specific parameters $\theta$, such as width and length, and returns a fixed-size vector $\vec x$ that captures the runtime parameters. The $\vec r_i, \vec r_q =$\txs{eval}($\vec x$,$\vec v$) function takes the device instance vector $\vec x$ and fixed-size voltage vector $\vec v$ capturing the nodal and branch voltages, and returns the fixed-size react vector $\vec r_q$ and fixed-size resist vector $\vec r_i$. The \txs{setup} function is invoked once for each unique PDK-instantiated device, \txs{init} is invoked once per device instance, and \txs{eval} is invoked multiple times per instance depending on the simulation.


\subsection{Python to Traceable JAX Functions}\label{sec:tech:jaxcompiler}

\tool{}'s \jax{} transpiler pass converts generated Python code to \jax{} Python code. \jax{} supports AD and just-in-time (\jit{}) compilation of numeric computing code, enabling GPU acceleration and parallelization of programs and optimization with gradient-based methods. \jax{} requires all data-dependent control flow (e.g., conditionals) to be implemented with equivalent \jax{} control flow (e.g., \txs{jax.lax.cond} calls) and prohibits arrays to be modified in place. More traceable variables in a function increase complexity in code generated by AD and \jax{} compilation, especially if the traced code contains many \txs{jax.lax.cond} calls.

\prh{Tracing Optimizations.} To reduce the complexity of the generated \jax{} code, \tool{}'s \jax{} transpiler pass uses a program analysis technique called \textit{taint analysis}. The transpiler takes a taint variable (TV) list which specifies the subset of PDK, device instance, and voltage input parameters with respect to which the user wishes to autodifferentiate. Then taint analysis conservatively identifies TV-dependent variables and conditional branches by iteratively propagating variable dependencies through a function AST. For the BSIM4 model to support AD, these identified operations must be traceable. All non-TV-dependent variables are resolved as runtime constants.

\prh{Conditionals.} The TV-dependent conditionals identified using the taint analysis technique are converted to \jax{} control flow, while all other conditionals are left as Python conditionals. This reduces unnecessary complexity while preserving the intended gradient computation paths.

\prh{Logic.} Scalar math operations are transformed into \txs{jax.numpy} operations to allow tracing. Bool casts become \txs{(x != 0)} while float casts are removed to prevent evaluating tracers without a defined value, as such an absence would throw errors. Integer literals become floating point literals to prevent \txs{int64} overflows. Integers used in discrete operations are preserved. Math operations with undefined behaviors (square root, log, division) are replaced by safe math operations that return finite surrogate values for undefined outputs, because in vectorized simulations of the model, both branches of a \txs{jax.lax.cond} call are computed and the unused branch may exercise undefined behavior in these operators.

\section{\circulax{} Integration and Device Library}\label{sec:tech:circsim}

The \tool{} transpiler produces device models compatible with \circulax{}, a  SPICE-like simulator built on \jax{} that supports AC, DC, and transient analysis of analog circuits. Section~\ref{sec:tech:devlib} describes \tool{}-generated \circulax{} device models and Section~\ref{sec:tech:pdklib} describes \tool{}-generated PDK libraries.

\subsection{\tool{}-Generated Differentiable Device Models}\label{sec:tech:devlib}

\circulax{} requires each device model to be defined as a function that accepts the input voltages $\vec v$ and the runtime parameters $\vec x$ as input. \tool{} uses \jax{}'s AD routine to trace through the function logic and derive a sensitivity function that computes the react and resist outputs with respect to $\vec x$ and $\vec v$.

When provided with a SPICE-like circuit description by the user, \circulax{} composes these device models to produce a circuit evaluation function, which is then invoked during simulation to produce a simulation result $\vec o$. Backpropagation is then used to compute $\partial \vec o/\partial x$ and $\partial \vec o/\partial \vec v$. \tool{} uses a combination of the chain rule and the sensitivity of the runtime parameters with respect to $\theta$ to compute $\partial \vec o/\partial \vec \theta$:

\begin{equation}
    \partial \vec{o}/\partial \vec{\theta}
    = \partial \vec{o}/\partial \vec{x}
    \cdot \partial \vec{x}/\partial \vec{\theta}
    \label{eq:chain_rule}
\end{equation}

The runtime parameter sensitivity $\partial \vec x/\partial \vec \theta$ is computed by tracing the \txs{init} function and auto-differentiating it. The \txs{init}($\phi$, $\theta$) function generated for \circulax{} combines the logic of the \txs{init} and \txs{setup} functions from the transpiler output and is only traceable with respect to $\theta$. With \tool{}'s chain rule extension, we can compute the sensitivity of the simulation result with respect to device parameters. 

\prh{Compatibility Optimizations.} \jax{} supports compilation of programs to XLA, \jax{}'s default compilation backend for acceleration. The circuit evaluation function is lowered to XLA. As a compatibility optimization, the \txs{init} function is autodifferentiated but not lowered to XLA, as it contains operators not supported by XLA. Because \jax{} does not support tracing through compositions of host-executed Python and XLA implementations, the \txs{init} and \txs{eval} sensitivities are individually computed and then composed with \eqref{eq:chain_rule}.

\prh{Performance Optimizations.} \circulax{} evaluates devices in parallel in a single batched operation. We modify \circulax{} to have the option of solving device physics sequentially to reduce the memory footprint during backpropagation. 

\prh{Shape Transformations.} The \jax{} transpiler reshapes the outputs of the \txs{eval} function into the format required by \circulax{}.

\subsection{PDK Support}\label{sec:tech:pdklib}

\tool{} automatically generates PDK libraries from the differentiable device models and a \textit{PDK specification}, a JSON file which describes the foundry-defined device parameters $\phi$ as a function of transistor width and length. The PDK specification is automatically extracted from the PDK card. \tool{} builds a PDK library that maps each PDK device model name to a specialized device map. The specialized device map translates bins, which define intervals of width and length for which the parameters apply, to a partially evaluated version of the device init function with $\phi$ populated. At each optimizer iteration, the appropriate \txs{init} function for each device is selected from its current width and length.

\subsection{Limitations}\label{sec:tech:limitations}

\tool{}'s transpiler does not support \txs{for} and \txs{while} loops. For each device model, all loops were bounded and so we hand-unrolled each loop. Loop unrolling is a standard compiler optimization, so this would be easy to automate.

\newcommand{\bopt}[0]{\txs{BOpt}}
\newcommand{\fdiff}[0]{\txs{FDiff}}
\newcommand{\adiff}[0]{\txs{ADiff}}

\section{Evaluation}

\subsection{Benchmark Circuits}\label{sec:eval:benchmark}

\begin{table}
\caption{Device, PDK, and Circuit Information}\label{tbl:methodology}
\begin{tabularx}{\linewidth}{|c|c|XXX|}
\hline
\multicolumn{5}{|c|}{BSIM 4 Device Model}\\
\hline
\multirow{2}{*}{Language}& \multirow{2}{*}{LoC}& \multicolumn{3}{|c|}{\# Parameters}\\
\cline{3-5}
 &  & Total & Discrete & Continuous\\
\hline
Verilog-A & 12654 & 899 & 33 & 866\\
Python & 46165 & 899 & 33 & 866\\
JAX &  58900 & 899 & 33 & 866\\
\hline
\hline
\multicolumn{5}{|c|}{SKY130 PDK ~$\bullet$~ \textit{all transistor models use BSIM4 device model}}\\
\hline
\multirow[c]{3}{*}{Transistor Model} &\multirow[c]{3}{*}{Used} & \multicolumn{3}{|c|}{\# Parameters}\\
\cline{3-5}
& & & Free & Free\\
& & Fixed & Discrete & Continuous\\
\hline
\txss{nfet\_01v8\_lvt}& $\checkmark$ & 885 & 2 & 9+3$^{\dagger}$\\
\txss{nfet\_01v8}& $\checkmark$ & 885 & 2 & 9+3$^{\dagger}$\\
\txss{nfet\_g5v0d10v5}&  & 885 & 2 & 9+3$^{\dagger}$\\
\txss{pfet\_01v8\_lvt}& & 885 & 2 & 9+3$^{\dagger}$\\
\txss{pfet\_01v8}& $\checkmark$ & 885 & 2 & 9+3$^{\dagger}$\\
\txss{pfet\_01v8\_hvt}& & 885 & 2 & 9+3$^{\dagger}$\\
\txss{pfet\_g5v0d10v5}& & 885& 2 & 9+3$^{\dagger}$\\
\hline
\end{tabularx}

\begin{tabularx}{\linewidth}{|c|c|cccc|}
\hline
\hline
\multicolumn{6}{|c|}{Benchmark Circuits ~$\bullet$~  \textit{all circuits use SKY130}}\\
\hline
Circuit  & \# Design  & \# Trans- & Res- & & \\
 &   Parameters & istors &\ istors &\ Caps & \ Switches\\
\hline
Track and Hold  & 1& 1& 0& 1& 0 \\
Bootstrap  & 9&1 & 0& 3& 5 \\
LDO  & 36& 6& 3& 2 & 0\\
\hline
\end{tabularx}
\par\vspace{2pt}
\noindent\parbox{\linewidth}{%
    \footnotesize $^{\dagger}$ Non-design parameters (\S\ref{sec:eval:benchmark}).
}
\par\vspace{2pt}
\end{table}

Table~\ref{tbl:methodology} presents the benchmark circuits, all implemented with the SKY130 PDK. The number of design parameters to optimize and the count of each device type are described in the table. Three benchmark circuits are evaluated -- a track and hold circuit, a bootstrapped sampling switch (bootstrap) circuit, and a low-dropout regulator (LDO). All designs use the PDK-provided transistor model and ideal models for passives and switches. The SKY130 PDK is an open-source process design kit which uses the level 54 BSIM4 device model. It exposes twelve continuous design parameters to the user, three of which are operating temperature, which we don't optimize, and the number of square units in drain and source diffusion areas which are hard-coded to zero. The underlying BSIM4 model is 12.7K lines of \veriloga{} code, and exposes 33 discrete parameters and 866 continuous parameters, all but 14 of which are set by the foundry.

\prh{Track and Hold.} Track and Hold is an analog circuit commonly used to hold voltages constant as they are measured for analog-to-digital converters.
This circuit is sourced from \cite{Ngspice_Notebook} and used as a sanity test in our analysis. In track mode, the output voltage follows an input voltage. In hold mode, the output holds the last voltage sampled in track mode. The operating mode is set by a clocked gate voltage on the transistor to control whether it conducts the input voltage across its drain-source terminals. The figure of merit (FoM) for this circuit minimizes the root mean-squared error (RMSE) of the circuit's transient voltage trajectory with respect to a reference trajectory while minimizing the width of the transistor. We optimize the width of the transistor, with an initial width of \SI{4}{\micro\meter}.



\prh{Bootstrap.} Bootstrap sampling switches are used to maintain an approximately constant VGS for a MOSFET regardless of the input signal, with the goal of minimizing sample distortion. Nine parameters are optimized for one transistor.
The benchmark circuit is sourced from~\cite{ICONS}. The FoM for this circuit aims to minimize transistor area while maintaining a third-harmonic distortion (HD3) below -75 dB. All nine transistor design parameters are optimized for this task. These FoM metrics are computed from a transient analysis. The specified SKY130 device minimum transistor length L = \SI{0.15}{\micro\meter} is used as an initial guess. An initial width W = \SI{60.75}{\micro\meter} is computed from an analytically derived W/L ratio of 405, which achieves HD3 less than -75 dB.


\prh{LDO.}
Low Dropout Regulators are voltage regulators used to maintain constant output voltage despite changes in input voltage or load current. This circuit is based on the SimpleLDO circuit from~\cite{AnalogGym}. Thirty-six parameters are optimized for four transistor groups. 
The FoM for this circuit jointly minimizes transistor area, minimizes the DC operating point bias, maximizes the PSRR which is equivalent to minimizing PSRR shortfall, and minimizes the quiescent power consumption. The PSRR is an AC metric that measures suppression of supply ripples in the input signals. These metrics are computed from a DC operating point analysis and AC analysis.
Since LDOs normally operate at varying loads and frequencies, the FoM accounts for multiple load current magnitudes and input signal frequencies. 

\subsection{Evaluation Methodology}

Each circuit's parameters are optimized with a loss function that captures the FoM for the circuit. For all optimization runs for a particular circuit, the same initial parameterization is used. For each optimizer, we set the maximum number of iterations to 40 iterations, except for \txs{BO-4X}, which uses 160 iterations. All optimization runs were performed on an x86-64 CPU machine with 16 GB of memory and an Intel Core Ultra 7 155H CPU. We \jit{} the circuit simulation for all optimization runs for improved performance.

\prh{Gradient-based Optimization.} To perform gradient-based optimization of circuit design parameters, the Adam optimizer is used with a fixed learning rate of 0.05, with all other parameters at default. A scaling term on parameter updates accounts for the varying scales of parameter values.

Two methods to calculate the gradients are used in the evaluation. We use gradients calculated by backpropagating through the automatically differentiated \tool{}-generated model (\txs{ADiff}). As a baseline, we also use finite difference methods (\txs{FDiff}) to approximate the gradient. The second-order central finite difference is used as it delivers good gradient approximations, and falls back to a first-order forward/backward finite difference on parameter boundaries.

\prh{Bayesian Optimization.} We evaluate gradient-based optimization with a black-box Bayesian optimization (\txs{BO-S} and \txs{BO-4X}) baseline, a gradient-free optimization method that uses a Gaussian process surrogate model to approximate the relationship between the parameters and the loss function. The \txs{BO-S} execution matches gradient-based methods with 40 optimization iterations and the \txs{BO-4X} execution uses $4 \times 40 = 160$ iterations. The \txs{BO-4X} baseline allocates excess solver iterations to account for higher-dimensional search spaces, giving Bayesian optimization a larger iteration budget to find a solution than gradient-based methods.



\prh{Metrics.} For each optimization approach, we report the discovered optimal loss and the circuit-specific metrics that comprise the FoM. To quantitatively evaluate the accuracy of the \tool{} model-generated parameter sensitivities relative to the finite difference baseline, we report the max peak normalized error (MPNE) between \txs{ADiff} and \txs{FDiff} gradients, which reports the highest error between them.

For evaluating performance, we report the optimization and compilation memory footprints in megabytes (MB). The optimization run time is broken down into the time required for each circuit solve (\txs{time/solve}), the time for each optimizer iteration (\txs{time/iter}), and the total optimization time (\txs{total}). We also report the number of circuit solves ($\# iter \times \# sims/iter$) for each optimizer.





\subsection{Analysis of Optimizer Results}

Tables~\ref{tbl:opt:trackandhold}, ~\ref{tbl:opt:bootstrap}, and ~\ref{tbl:opt:ldo} report each optimizer's parameterizations for each benchmark circuit. 
The gradient-based methods find better parameterizations than \txs{BO-S} and \txs{BO-4X} as the number of design parameters increases. The \tool{}-generated \txs{ADiff} flow produces nearly equivalent parameterizations to \txs{FDiff} in far fewer circuit solves. For Track and Hold (Table~\ref{tbl:opt:trackandhold}) ADiff converges in 66.7\%  fewer solves than FDiff and achieves worse loss but smaller width than \txs{BO-S} and \txs{BO-4X}. In Bootstrap (Table~\ref{tbl:opt:bootstrap}), \txs{ADiff} finds an optimum in 9.3\% of the circuit solves of FDiff and finds parametrizations that have a smaller loss, smaller area, and satisfactory HD3 compared to \txs{BO-S} and \txs{BO-4X}. For the LDO (Table~\ref{tbl:opt:ldo}), \txs{ADiff} finds an optimum in 2.4\% of the circuit solves of \txs{FDiff} while achieving a 40\% lower loss, lower PSRR shortfall, and lower DC operating point bias than \txs{BO-S} and \txs{BO-4X}. In general, Bayesian optimization-based methods have increased difficulty finding good parameterizations as the search space expands.

\begin{table}[t]
\footnotesize
\caption{Track and Hold Circuit Width Parameter and Figure of Merit}
\label{tbl:opt:trackandhold}
\begin{tabularx}{\linewidth}{|l|X|cX|c|}
\hline
Method & W ($\mu m$)& \# Solves & Loss & RMSE\\
\hline
BO-S & 4.879  & 41 & $2.786 \cdot 10^{-7}$&  $5.278 \cdot 10^{-4}$\\
BO-4X & 4.879  & 161 & $2.786 \cdot 10^{-7}$&  $5.278 \cdot 10^{-4}$\\
FDiff & 4.430  & 42& $3.862 \cdot 10^{-6}$&  $1.965 \cdot 10^{-3}$\\
ADiff & 4.430  & 14& $3.826 \cdot 10^{-6}$& $1.965 \cdot 10^{-3}$\\
\hline
\end{tabularx}
\end{table}

\begin{table}
\caption{Bootstrap Circuit Parameters and Figure of Merit}
\label{tbl:opt:bootstrap}
\footnotesize
\begin{tabularx}{\linewidth}{|l|XXccccccc|}
\hline
Method & W ($\mu m$) & L ($\mu m$) & ad & as & pd & ps & sb & sa & sd \\
\hline
BO-S & 60.75& 0.15& 0& 0& 0& 0& 0 & 0& 0 \\
BO-4X & 49.751 & 0.15 & 0& 0& 0 & 0 & 0 & 0& 0 \\
FDiff & 44.744& 0.15& 0& 0& 0 & 0 & 0 & 0& 0 \\
ADiff & 44.743& 0.15& 0 & 0& 0 & 0 & 0 & 0& 0 \\
\end{tabularx}

\begin{tabularx}{\linewidth}{|l|X|cXX|}
\hline
\hline
 &\#  &  & transistor  & harmonic  \\
Method & solves & loss &  area ($\mu m^2$) & distortion (db) \\
\hline
BO-S &  41 & 0.152 & 9.113  & -81.598 \\
BO-4X & 161 & 0.124 & 7.463 & -79.149 \\
FDiff &  441 & 0.112& 6.712  &-77.757 \\
ADiff &  41 & 0.112& 6.711  & -77.757 \\
\hline
\end{tabularx}
\end{table}

\begin{table}
\caption{Low Dropout Regulator}
\label{tbl:opt:ldo}
\footnotesize

\begin{tabularx}{\linewidth}{|l|cc|XXXX|}
\hline
 &  &  &   & max  &  &  DC Op. \\
 &  \#  &  & transistor  & quiescent  &  &  Point \\
Method &  solves & loss &  area ($\mu m^2$) & power (mW) & PSRR Shortfall (dB) &   Bias (mV)  \\

\hline
BO-S & 41& 5.163& 976.818 & 0.0813 & 3.198& 10.694  \\
BO-4X & 161 & 5.163 & 976.818 & 0.0813 & 3.198& 10.694  \\
FDiff & 1677& 3.060 & 1028.918 & 0.0838 & 2.070& 10.342 \\
ADiff & 41& 3.076& 1034.709 & 0.0840 & 2.066& 10.361  \\
\hline
\end{tabularx}%
\end{table}

\subsection{Sensitivity Analysis}

We next compare the agreement between gradients for \txs{ADiff} and \txs{FDiff}. We find \txs{ADiff} and \txs{FDiff} gradients agree across benchmark circuits, validating its accuracy in transient, AC, and DC analysis. For Track and Hold, the MPNE for output voltage gradients $\partial V_{out}/\partial p$ over all parameters $p$ is under $10^{-4}$ during transient analysis. For Bootstrap, the MPNE for HD3 gradients $\partial HD3/\partial p$ and $V_{GS}$ gradients $\partial V_{GS}/\partial p$ during transient simulation is under $10^{-3}$ and $0.15$, respectively, for all parameters. For the LDO, MPNE for PSRR gradients $\partial PSRR/\partial p$ is under $10^{-2}$ during AC analysis, and the MPNE for voltage gradients $\partial V_{out}/\partial{p}$ is under $10^{-5}$ for all parameters. To summarize, the gradient discrepancy between \txs{ADiff} and \txs{FDiff} is quite low, even in the worst case.

\begin{figure}
    \begin{subfigure}[b]{1.0\linewidth}
        \centering
        \includegraphics[width=1\linewidth]{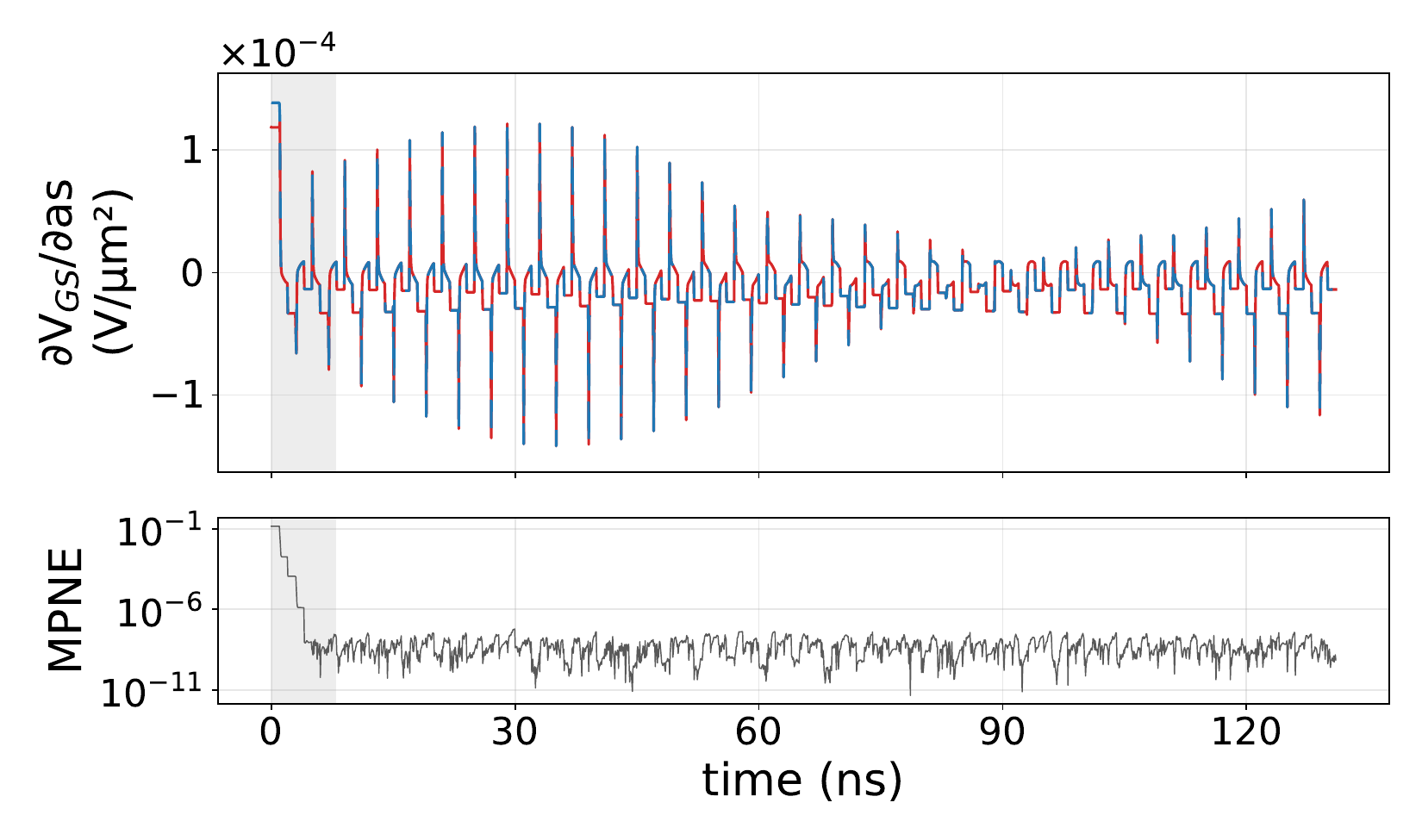} 
        \caption{$\partial V_{gs}/\partial as$ for Bootstrap Circuit}
        \label{fig:sens:bootstrap}
    \end{subfigure}
    
    \vspace{1.5em} 
    
    \begin{subfigure}[b]{1.0\linewidth}
        \centering
        \includegraphics[width=1\linewidth]{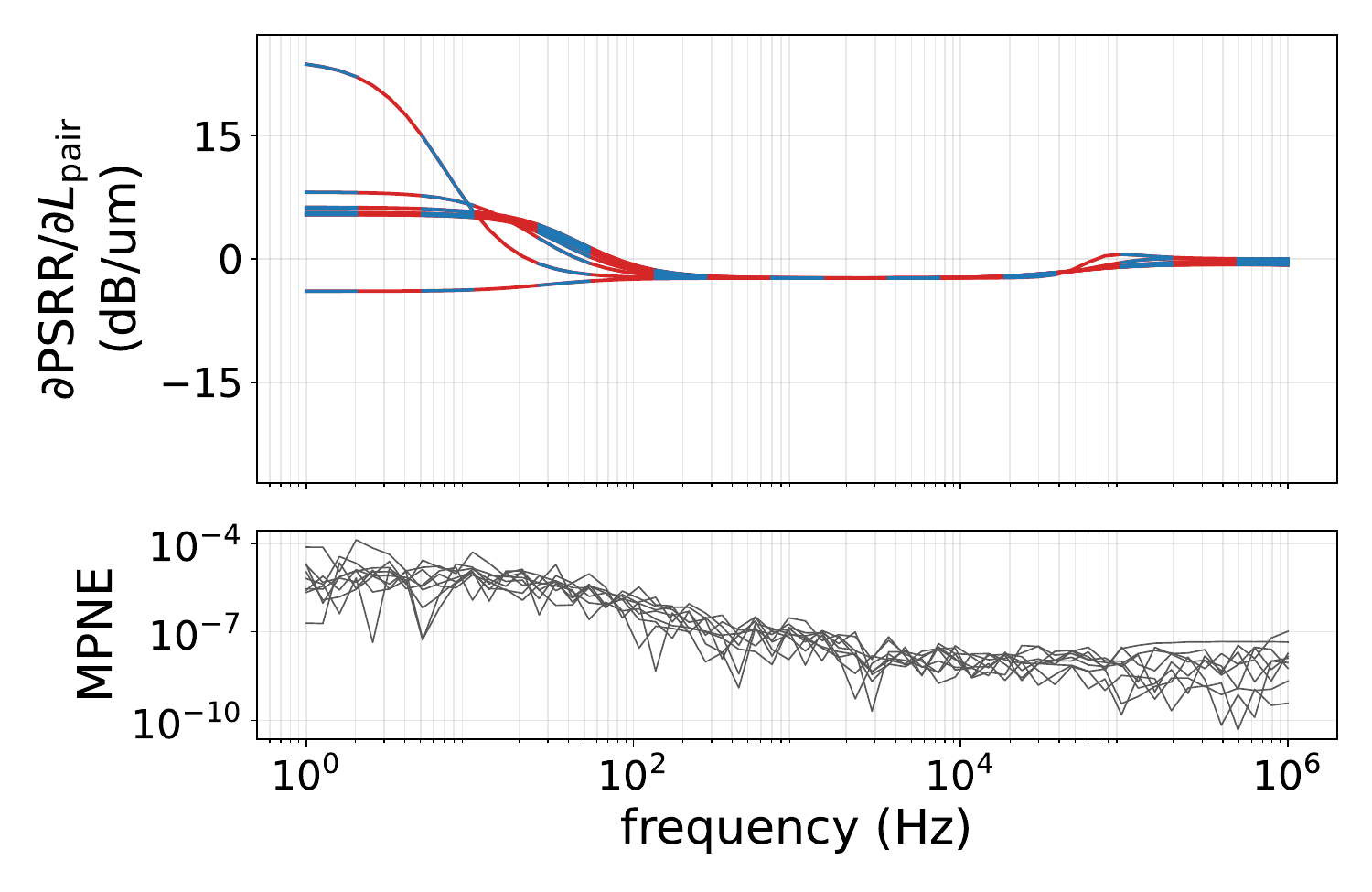} 
        \caption{$\partial PSRR/\partial L_{pair}$ for LDO}
        \label{fig:sens:ldo}
    \end{subfigure}
    
    \caption{Parameter Sensitivities for Bootstrap and LDO. \swatch{blue} is \txs{ADiff}, \swatch{red} is \txs{FDiff}, \swatch{black} is MPNE.}
    \label{fig:sens}
\end{figure}

Figures~\ref{fig:sens:bootstrap} and ~\ref{fig:sens:ldo} qualitatively present the gradients and MPNE for $\partial V_{GS}/\partial as$ and $\partial PSRR/\partial {L_{pair}}$.
Each figure displays the \txs{ADiff} and \txs{FDiff} computed gradients on top and the MPNE on the bottom. The x-axis is shared across stacked plots. Largely, the trajectories qualitatively agree. We note in the bootstrap $V_{GS}$ plot, \txs{ADiff} and \txs{FDiff} initially disagree -- in this instance \txs{FDiff} is reporting the incorrect gradient because its difference of the initial DC voltage at $t=0$ is very small such that the DC error meaningfully impacts its gradient. In contrast, \txs{ADiff} takes the instantaneous derivative. The MPNE of $\partial V_{GS}/\partial p$ in the quantitative analysis reflects this discrepancy. With this initial time segment excluded, the MPNE is under $10^{-3}$.



\begin{table}[t]
\footnotesize
\caption{Performance Analysis}
\label{tbl:performance}
\begin{tabularx}{\linewidth}{|c|cc|cXc|}
\hline
Method & \multicolumn{5}{|c|}{Track and Hold (\textit{transient (.TRAN) analysis, 2000 time steps})} \\
\hline
& \multicolumn{2}{|c|}{\textit{Memory (MB)}} & \multicolumn{3}{|c|}{\textit{Run time (s)}}\\
\hline
& optimization & compile & time/solve & time/iter & total \\
\hline
BO-S & 52  & 703  & 12.08 & 12.38 & 495.24 \\
BO-4X & 60 & 645 & 13.21 & 13.29 & 2126.94 \\
FDiff &  22  & 711  & 12.59 & 37.77& 528.85\\
ADiff & 153  & 5555  & 159.92 & 159.92 & 2238.84\\
\hline
\end{tabularx}

\begin{tabularx}{\linewidth}{|c|cc|cXc|}
\hline
Method & \multicolumn{5}{|c|}{Bootstrap \textit{(transient (.TRAN) analysis, 2000 time steps)}}  \\
\hline
& \multicolumn{2}{|c|}{\textit{Memory (MB)}} & \multicolumn{3}{|c|}{\textit{Run time (s)}}\\
\hline
& optimization & compile & time/solve & time/iter & total \\
\hline
BO-S & 25 & 743  & 25.58 & 26.22 & 1048.6  \\
BO-4X & 65 & 683 & 31.08 & 31.27 & 5003.91 \\
FDiff &  8  & 732  & 27.00 & 297.63 & 11905.30 \\
ADiff &  193 & 5576 & 355.98 & 364.88 & 14595.37 \\
\hline
\end{tabularx}
\begin{tabularx}{\linewidth}{|c|cc|cXc|}
\hline
Method &  \multicolumn{5}{c|}{LDO (\textit{DC (.OP) and AC Small Signal (.AC) analysis})} \\
\hline
& \multicolumn{2}{|c|}{\textit{Memory (MB)}} & \multicolumn{3}{|c|}{\textit{Run time (s)}}\\
\hline
& optimization & compile & time/solve & time/iter & total \\
\hline
BO-S & 22 & 338 & 8.40 & 8.60 & 344.35 \\
BO-4X & 231 & 333 & 16.74 & 16.84 & 2695.19 \\ 
FDiff &  35 & 329  & 0.13 & 5.35  & 214.01 \\
ADiff &  201 & 5014 & 34.92  & 35.80 & 1431.91 \\
\hline
\end{tabularx}

\end{table}

\subsection{Performance Analysis.}

Table~\ref{tbl:performance} presents the run time and memory footprint breakdown for each optimizer. At compile-time, \txs{ADiff} requires more memory than the baselines due to the AD pass, which is required to directly compute gradients. During optimization, \txs{ADiff} requires more time for each circuit solve compared to baseline methods due to the backward pass. However, it requires fewer circuit solves for each iteration than \txs{FDiff}, so the time per iteration and total run time are closer to \txs{FDiff} depending on the benchmark -- especially if transient simulation is performed. In cases where AC analysis is performed, such as with the LDO, \txs{ADiff} solve times are significantly higher than the non-autodifferentiated baselines due to mixed second-order derivatives computed in the backward pass of AC systems, while the forward-only pass is extremely cheap to compute. 

We observe that although \txs{ADiff} has higher memory and run time requirements, these performance figures do not drastically increase with increasing number of design parameters. Comparing Track and Hold ($N=1$) and Bootstrap ($N=9$) under an identical 2,000-step transient setup shows that \txs{ADiff}'s per-solve run time and optimization memory grow by only $2.2\times$ and $1.3\times$, respectively, despite a $9\times$ increase in parameters. In contrast, \txs{FDiff's} iteration time increases by 8x. 

There are several reasons for this performance difference. We note that \circulax{}, which implements the backward pass, is about a year old and primarily builds circuit simulation algorithms using generic optimization and simulation libraries for \jax{}, such as \txs{optimistix} and \txs{diffrax}. Therefore, \circulax{} takes little advantage of any domain-specific characteristics of circuit simulation algorithms. \tool{} also does not yet apply performance optimizations when transpiling the model and relies on \jax{} to discover optimization opportunities. With further development in both libraries, the compile and solve times could decrease.

\section{Related Work}

\prh{Methods for Computing Circuit Gradients.} 
The ``adjoint sensitivity analysis” technique in analog design automation, which involves building an auxiliary adjoint circuit to compute sensitivities, has since been studied extensively~\cite{director1969generalized, visweswariah2000noise, hu2020adjoint, li2023circuit}. These works focus on device-level optimization, specializing in circuit-specific metrics such as delay or gain, require linearization of the circuit components, and typically are limited to simple transistor models. In contrast, \tool{} is able to take gradients through circuits that use significantly more complex transistor models via AD; our experiments involve transistor models based on BSIM's level 54 transistor model. Finite difference methods are also used to compute gradients without requiring AD through the circuit model. For an $N$ parameter gradient, forward-backward finite difference methods require $N+1$ simulations but induce higher gradient error, and central methods require $2N$ simulations and have tighter gradient error. \tool{} enables direct computation of gradients from a single circuit simulation by backpropagating through the simulation's numeric computation. 


\prh{Gradient-based Circuit Optimization Tools.}

Circuit simulators such as Xyce and Circulax offer basic support for AD of device models and circuit simulations~\cite{xyce, circulax}. Xyce supports direct and adjoint sensitivities for device models, but only supports analytic derivatives for hand-differentiated linear devices and models generated by its ADMS back end. Other models, e.g., BSIM4 and BSIM-SOI, fall back to finite difference methods, which have higher gradient error and is simulation-intensive~\cite{xyce}. Additionally, Xyce falls back to numerical approximations for derivatives beyond first-order such as those associated with AD of AC simulation. In contrast, \tool{}'s transpiler enables compilation of device models directly to \jax{}-compatible Python, therefore enabling end-to-end AD of models to compute higher-order parameter sensitivities and GPU acceleration.

Moreover, Xyce models do not have Python representations while \tool{}, our transpiler, builds on general AD and \tool{}-compiled models compose directly with Python optimization and learning frameworks.

Production circuit optimizers such as WiCkeD use finite difference methods to approximate gradients in conjunction with a number of algorithmic optimization techniques~\cite{bHurmen2015wicked}. In principle, \tool{}'s differentiation method can be used in place of the described finite difference techniques.


\prh{Analog Circuit Optimization}
State-of-the-art analog circuit optimization for transistor sizing usually treats simulators as gradient-free black boxes, using Bayesian optimization~\cite{lyu2018batch}, reinforcement learning~\cite{wang2020gcn,settaluri2020autockt}, or language models~\cite{yin2024adollm}. Recent work combining Bayesian optimization with circuit parameter gradients improved sampling efficiency~\cite{lyu2024bayesianwithsensitivity}. However, their framework uses adjoint sensitivity analysis and is therefore limited to circuits composed of simple transistor models. \tool{} complements this optimization approach by providing accurate gradients for complex device models, e.g., BSIM4, enabling hybrid optimization of more complex circuits.

\section{Conclusion}

We present \tool{}, a \veriloga{}-to-\jax{} transpiler that enables automatic differentiation of compact device models frequently used in PDKs, enabling end-to-end gradient-based optimization of analog circuits within Python simulation and machine learning ecosystems. Because \tool{} enables direct derivation of parameter sensitivity functions from the device model through tracing, it avoids the $O(N)$ simulation cost of finite-difference approximations by computing exact design-parameter sensitivities from a single circuit simulation. We demonstrate that \tool{}-generated models can be successfully integrated into \circulax{} to optimize SKY130 circuits with far fewer circuit solves than finite difference methods. Furthermore, we demonstrate that gradient-based optimization enabled by \tool{} can find better optima in multi-dimensional spaces around an initial point than Bayesian optimization, offering the potential of an efficient hybrid method. With further algorithmic and performance optimizations, we believe that the absolute runtime and memory footprint can be further improved. 

\clearpage

\bibliographystyle{IEEEtran}
\bibliography{references}     

\end{document}